\documentclass{pasj00}

\begin{document}
\SetRunningHead{KASE AND MAKINO}{SUBHALO DISTRIBUTION}
\Received{2006/03/01}
\Accepted{2000/00/00}

\title{Missing Dwarf  Problem in Galaxy Clusters}

\author{Hiroyuki \textsc{Kase}}%
\email{hkase@astron.s.u-tokyo.ac.jp}
\author{Junichiro \textsc{Makino}}
\affil{Department of Astronomy, School of Science, The University of
Tokyo,\\7-3-1 Hongo, Bunkyo-ku, Tokyo 113-0033}
\and
\author{Yoko \textsc{Funato}}
\affil{Department of General Systems Studies, College of Arts and Sciences,  The University of
Tokyo,\\3-8-1 Komaba, Meguro-ku, Tokyo 153-8902}



\maketitle

\begin{abstract}
We investigated the formation and evolution of CDM subhalos in
galaxy-sized and cluster-sized halos by means of $N$-body simulations.
Our aim is to make clear what the ``dwarf galaxy problem'' is.  It has
been argued that the number of subhalos in simulated galaxy-sized halos is too
large compared to the observed number of dwarfs in the local group,
while that in cluster-sized halos is consistent with observed number
of galaxies in clusters such as the Virgo cluster.  We simulated nine
halos with several different mass resolutions and physical scales.
We found that the dependence of the cumulative number of subhalos
$N_c$ on their maximum circular velocity $V_c$ is given by $N_c\propto
V_c^{-3}$, down to the reliability limit,  independent of
the mass of the main halo. This implies that simulations for cluster-sized
halos give too many halos with $V_c \sim 140{\rm km/s}$ or less.
Previous comparisons of cluster-sized halos gave much smaller
number of subhalos in this regime simply because of their limited
resolution.  Our result implies that any theory which attempts to
resolve the missing dwarf problem should also explain the discrepancy
of the simulation and observation in cluster-sized halos.
\end{abstract}

\section{Introduction}

The Cold Dark Matter(CDM)
scenario\citep{1978MNRAS.183..341W} has been the standard theory of
the formation and evolution of the structures in the Universe.   In
this scenario, galaxies and clusters of galaxies are formed bottom-up.
It has been remarkably successful in explaining the large
scale structures \citep{1985ApJ...292..371D} and numerous
observational results (e.g. \cite{2005Natur.435..629S}). 
Until
recently, however, the small-scale  structures of CDM, like subhalos
in a galaxy-sized halo, could not be studied by numerical simulation
because of the lack of the computational power.  Recent improvement of
computational powers made it possible to study not only dark matter
halos, but subhalos in a parent 
halo.  Usually these subhalos are interpreted as corresponding
to galaxies in cluster-sized halos and satellite dwarf
galaxies in galaxy-sized halos.

\citet{1999ApJ...522...82K} and \authorcite{1999ApJ...524L..19M} (1999,
hereafter M99) reported that  about 1000 subhalos formed  in a
simulated galaxy-sized  halo. The number distribution of subhalos as
the function of the circular velocity normalized by that of the parent
halo turned out to be remarkably similar for galaxy-sized and
cluster-sized halos. This result, of course, is the direct outcome of
the scale-free nature of the gravity and almost power-law amplitude
of the density fluctuation.

This similarity, which looks quite natural from the theoretical side, showed
serious discrepancy with observation. The number of subhalos in the
simulated CDM halo agreed well with the observation of the cluster of
galaxies, while the simulation of a galaxy-sized halo predicted too
many dwarf-sized halos.  This discrepancy is now known as the ``dwarf
galaxy problem'', or ``missing-dwarf problem'', and has been one of
the main topics in the study of galaxy formation and structure
formation in the Universe.

Solutions so far proposed for this problem can be classified into the
following two categories:

\begin{itemize}
\item Dark matter is different from the CDM model in small scales.
\item Only a small fraction of subhalos is observed as dwarf galaxies.
\end{itemize}

Many modified dark matter models had been proposed. Self Interacting
Dark Matter\citep{2000PhRvL..84.3760S} is a CDM with small
self-interaction cross section, which dumps small-scale fluctuation.
Warm Dark Matter\citep{1986ApJ...304...15B} has a power spectrum
which has cut off at small scale. Some simulations with these dark
matter models have been done. (e.g. \cite{2001ApJ...547..574D},
\cite{2000ApJ...542..622C})

If we assume that the standard CDM is correct and consider the
possibility that not all
subhalos are observed as dwarf galaxies, the question is which subhalos
are observed as dwarf galaxies.  \citet{2002MNRAS.335L..84S} argued that
observed dwarf galaxies are presently most massive subhalos, by
comparing numerical result with observation of velocity dispersion
profile of dwarf galaxies.  On the other hand,
\authorcite{2004ApJ...609..482K} (2004, hereafter K04) tracked each
subhalo's evolution and showed many low-mass subhalos were massive in
the past.  Additionally, they introduced the model of star formation
condition, and claimed that the number and distribution of satellite
dSph galaxies was in agreement with that of such previously massive
subhalos.

Before jumping to such a solution, however, we should understand what
is really the problem.  In previous studies, in order to determine the
number of subhalos with circular velocity $\sim10\mathrm{km/s}$, the detection
limit was set to a few tens of particles. It is not clear whether or not
we can rely on the result obtained with such a small number of
particles.  Both softening parameters and the two-body relaxation might
significantly affect mass and density profiles of subhalos
(e.g. \cite{1996ApJ...457..455M},
\cite{2004MNRAS.348..977D}).
\citet{2000ApJ...544..616G} showed that the low mass end of the circular
velocity distribution of subhalos is affected by numerical resolution
by comparing the results of two cluster-scale simulations with
different resolutions.  Same tendency can be seen in the more
recent studies  (e.g. \cite{2004MNRAS.352..535D},
\cite{2004MNRAS.355..819G} and \cite{2005MNRAS.359.1537R}).
This means that the agreement in a cluster scale described in M99
might be a numerical artifact.  

In the present paper, as a first step to understand the formation and
evolution of subhalos, we investigated the reliability of size
distribution of subhalos  in simulation results.  We performed  large-scale
$N$-body simulations of formation of CDM halos.  We simulated
both galaxy-sized and cluster-sized  halos.

We carried out $N$-body simulations of nine halos of different masses
and resolutions and analyzed the distribution of subhalos.  From the
results of these simulations, we estimated the effect of the mass
resolution on the distribution of subhalos. When the number of
particles in one subhalo is small, both the mass and the circular
velocity of that halo are affected, resulting in the deviation from
simple power-law dependence.
We formulated the reliability criterion for mass and
circular velocity.

We then compared the velocity distribution function obtained by
$N$-body simulations with that of observed galaxies in the Virgo
Cluster. We found that the observational result is in good agreement
with our medium-resolution result, which is affected by small-$N$
effect around $V_c \sim 100\mathrm{km/s}$.
In other words, our high-resolution simulation
predicts the overabundance of moderate-sized galaxies.

Previously the mass function of galaxies in a cluster has been
believed to be well reproduced by CDM simulations (e.g. M99).
Our result shows that the agreement, in the range of $V_c=140\mathrm{km/s}$,
is due to the lack of mass resolution in previous
simulations. The discrepancy between the simulated halos and observed
galaxies extends to higher mass than in the galaxy scale.  This is
simply because there was no satellite of that mass in galaxy-scale
calculations. 
Our result
indicates that an analogue to the ``missing dwarf problem'' exists
also in clusters of galaxies.  Therefore, the ``solutions'' for the
missing dwarf problem should give a distribution consistent with observations
in both the galaxy scale and the cluster scale.

The contents of this paper is as follows. In section 2, we describe
the models and method used in our simulations. We give detailed
explanation of the subhalo detection method we developed.  In section
3, we discuss the reliability limit of the number count of subhalos.
Finally we summarize our result and discuss possible solutions to the
missing dwarf problem  in section 4.


\section{Model and Method}
\subsection{Cosmological $N$-Body simulation}
\subsubsection{Initial condition}

We adopted the Standard CDM model(SCDM), so that we can directly
compare our result with that of M99 who adopted SCDM. Cosmological
parameters are $\Omega_M=1.0$, $\Omega_\Lambda=0.0$, $H_0=50\mathrm{km/s}$,
$\sigma_8=0.7$.  To generate the initial condition, we used the
standard re-simulation method\citep{1996ApJ...462..563N}. First we
simulated 15 comoving $\mathrm{Mpc}$ radius using $1.1\times 10^6$
particles($m=8.9\times 10^8\MO$) and 150 comoving $\mathrm{Mpc}$ radius
using $1.1\times 10^6$ particles ($m=8.9\times 10^{11}\MO$) to
select candidates for galactic and cluster halos from the distribution
of particles at $z=0$. To select halo candidates, we used the standard
Friends of Friends method (FoF, \cite{1985ApJ...292..371D}) with the
linking length which is 0.2 times the mean interparticle distance.
Then we selected particles inside the radius
four times larger than radius enclosing particles detected by FoF as
the region of high resolution calculation, and traced these particles
back to the initial condition.  We replaced these particles with high
resolution particles and re-ran the simulation. In this way, the
external tidal force from outside  the high resolution region was
correctly taken into account.

To generate initial conditions, we used the grafic2 package
\citep{2001ApJS..137....1B}. The calculation was done on an IBM pSeries690
of the Data Reservoir
Project\footnote{http://data-reservoir.adm.s.u-tokyo.ac.jp/} of the
University of Tokyo, and on a workstation with AMD Opteron 242 processors
and 16GB memory.

We prepared initial conditions of one region with three different
resolutions for both galaxy-scale and cluster-scale runs. These runs are
named C1-H, C1-M, and C1-L and G1-H, G1-M, G1-L, for cluster scales
and galaxy scales. Here, postfixes H, L and M denotes high, medium and
low-resolution runs, respectively. We also selected two regions
for additional cluster-scale runs with medium resolution. These runs
are named C2-M and C3-M. The initial redshift $z_{start}$ and
particle mass $m$  are listed in table \ref{tab:param}.

\begin{table*}[tb]
\begin{center}
\caption{Parameters for calculations.
}
\label{tab:param}
\begin{tabular}{c||ccccc}
\hline
run & $m$($\MO$)  & $\Delta t_{fin}$($10^6 \mathrm{yr}$) & $\epsilon_{fin}$($\mathrm{kpc}$) & $z_{start}$ & $z_{crit}$\\
\hline\hline
G1-L & $4.1\times10^6$ & $4.3 $ & $0.6$ & $63.1$ & $11.8$\\
\hline
G1-M & $1.2\times10^6$ & $2.9$ & $0.4$ & $69.4$ & $13.1$\\
\hline
G1-H & $5.1\times10^5$ & $2.2$ & $0.3$ & $74.0$ & $14.0$\\
\hline\hline
C1-L,C2-L & $4.1\times10^9$ & $6.5$ & $6$ & $31.8$ & $5.56$\\
\hline
C1-M,C2-M & $1.2\times10^9$ & $4.3$ & $4$ & $36.7$ & $6.54$\\
\hline
C1-H,C2-H & $5.1\times10^8$ & $3.2$ & $3$ & $40.3$ & $7.27$\\
\hline
\end{tabular}
\end{center}
\end{table*}

\begin{table*}[tb]
\begin{center}
\caption{Properties of halos.}
\label{tab:prop}
\begin{tabular}{c||ccccc}
\hline
run & $M_{200}$($\MO$) & $N_{200}$ & $R_{200}$($\mathrm{kpc}$) & $V_c(\mathrm{km/s})$ & $N_{sub}$\\
\hline\hline
G1-L & $2.3\times10^{12}$ & $570490$ & $344$ & $236$ & $220$\\
\hline
G1-M & $2.2\times10^{12}$ & $1768684$ & $334$ & $235$ & $601$\\
\hline
G1-H & $2.1\times10^{12}$ & $4179435$ & $334$ & $233$ & $1346$\\
\hline\hline
C1-L & $1.7\times10^{15}$ & $405729$ & $3067$ & $1703$ & $350$\\
\hline
C1-M & $1.4\times10^{15}$ & $1198496$ & $2934$ & $1495$ & $977$\\
\hline
C1-H & $1.6\times10^{15}$ & $3170198$ & $3044$ & $1669$ & $2019$\\
\hline\hline
C2-L & $8.2\times10^{14}$ & $199993$ & $2423$ & $1397$ & $129$\\
\hline
C2-M & $8.1\times10^{14}$ & $673493$ & $2416$ & $1373$ & $428$\\
\hline
C2-H & $8.2\times10^{14}$ & $1600368$ & $2424$ & $1203$ & $1010$\\
\hline
\end{tabular}
\end{center}
\end{table*}

\subsubsection{Time integration method}

The time integration was done in physical time and physical
coordinates.  In high-$z$ regime, we varied time step $\Delta t$
and the softening length $\epsilon$ to reduce the time integration
error.
We adopted the procedure similar to that used in
\citet{2004ApJS..151...13K}.  We used the following formulation
\begin{eqnarray}
\Delta t(z) =& \left(\frac{z_{crit}+1}{z+1}\right)^\frac{3}{2} \Delta t_{fin},\\
\epsilon(z) =& \frac{z_{crit}+1}{z+1} \epsilon_{fin},
\end{eqnarray}

where $z_{crit}$ is the redshift at which we switch from $z$-dependent
timestep and softening to constant timestep and softening. The
constant values are $\Delta t_{fin}$ and $\epsilon_{fin}$.  These
equations imply $\Delta t \propto \sqrt{1/\bar{\rho}}$ and $\epsilon
\propto a$ for $z > z_{crit}$, where $\bar{\rho}$ is the mean density,
$a$ is the scale factor.  The initial redshift($z_{start}$) and
$z_{crit}$ correspond to the standard deviation of density fluctuation
$(\rho-\bar{\rho})/\bar{\rho}$ is 0.2 and 1.0, respectively.
We chose $\Delta t_{fin}$ so that the energy conservation is better
than $3\%$ when calculations are done only for the high resolution
particles.  We chose $\epsilon_{fin}$ so that it is sufficiently small
compared to the size of subhalos. For runs with different $N$, we
varied $\epsilon_{fin}$ in proportion to $N^{\frac{1}{3}}$.  These
values are summarized in table \ref{tab:param}.

For actual time integration, we used a parallel implementation of
Barnes-Hut tree algorithm 
\citep{1986Natur.324..446B,2004PASJ...56..521M} on
GRAPE-6\citep{2003PASJ...55.1163M} and GRAPE-6A
\citep{2005PASJ...57.1009F}. The opening angle was $0.3$ for all run.
Time integration was done using the standard leapfrog.

\begin{figure*}[t]
\begin{center}
\FigureFile(0.45\hsize,0.45\hsize){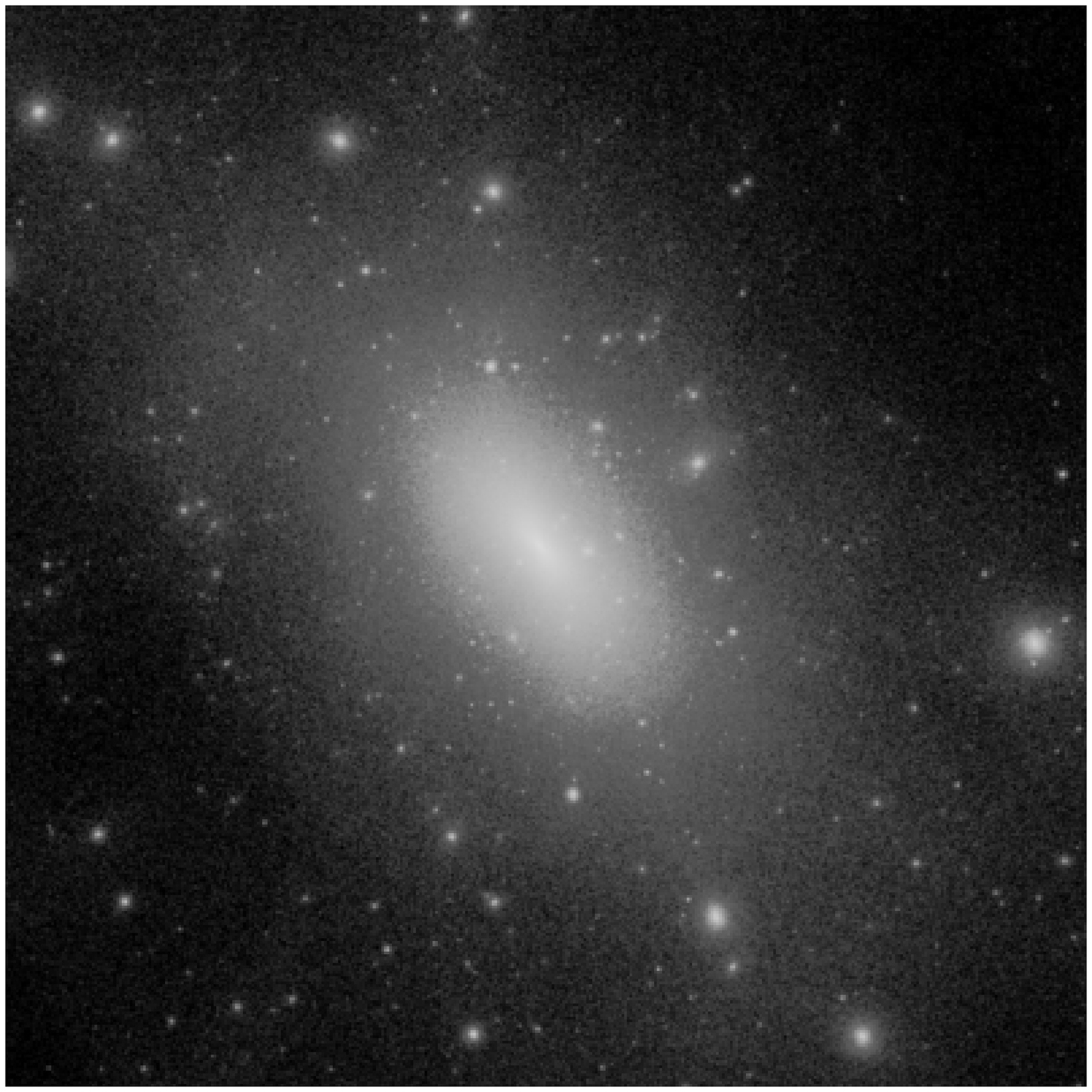}
\FigureFile(0.45\hsize,0.45\hsize){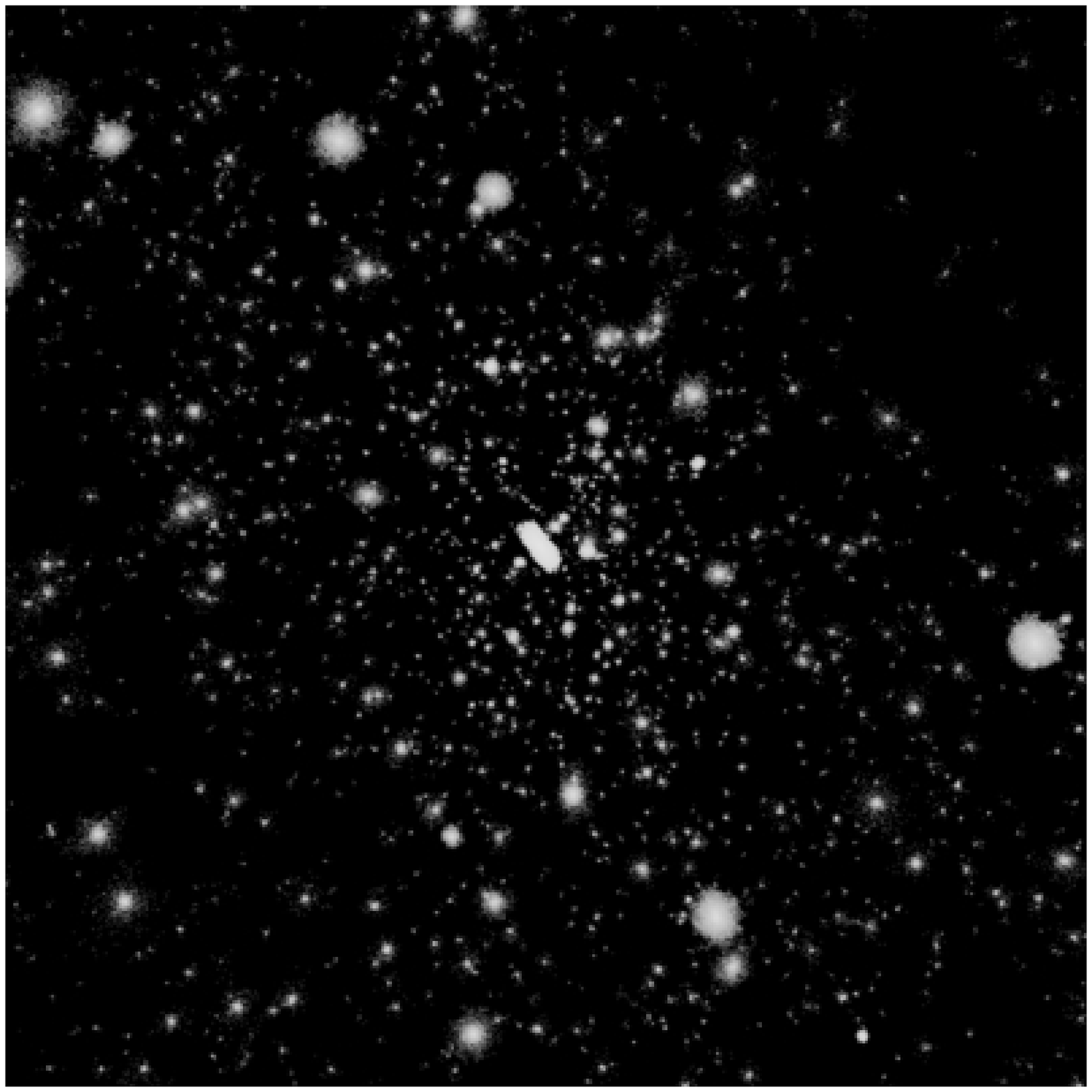}\\
\FigureFile(0.45\hsize,0.45\hsize){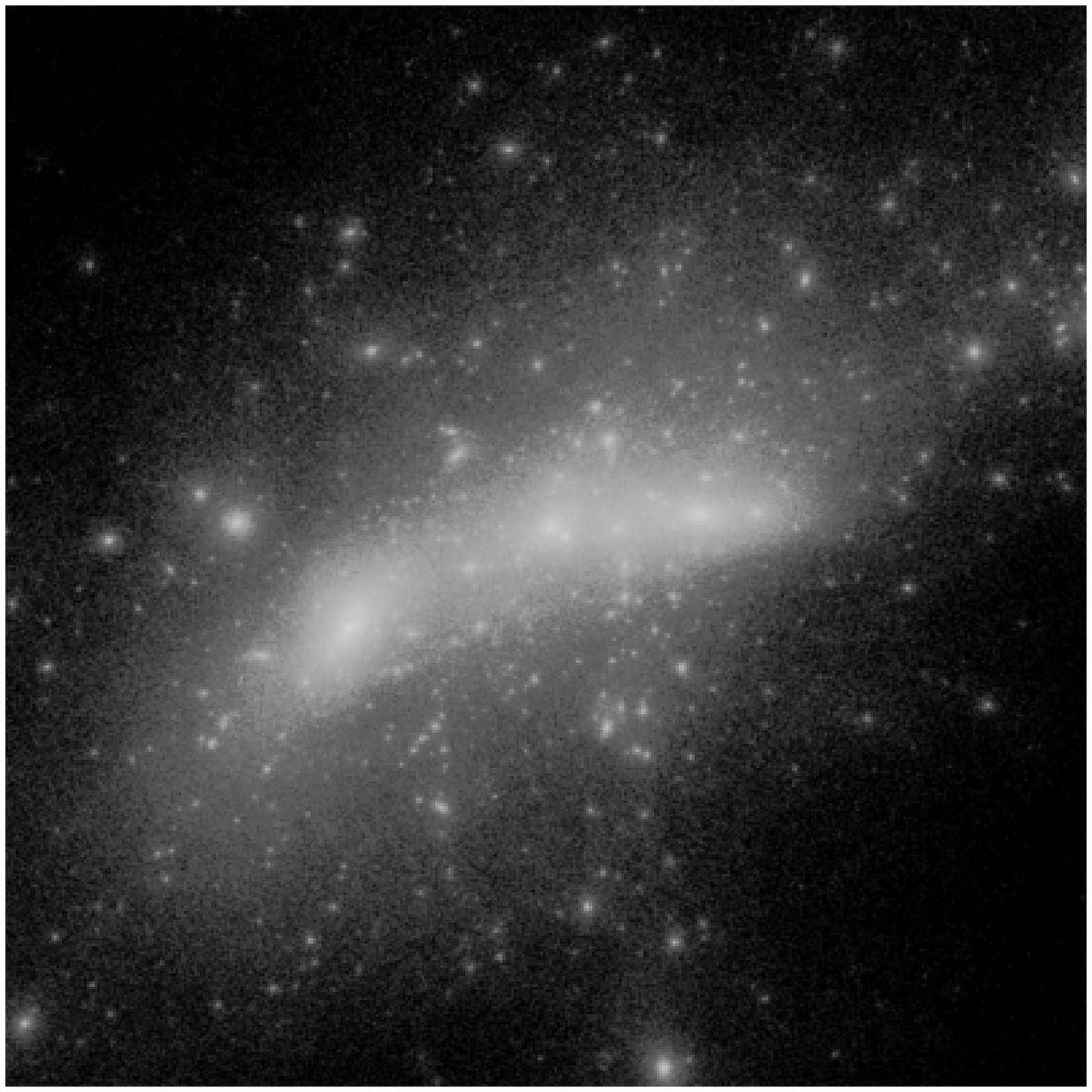}
\FigureFile(0.45\hsize,0.45\hsize){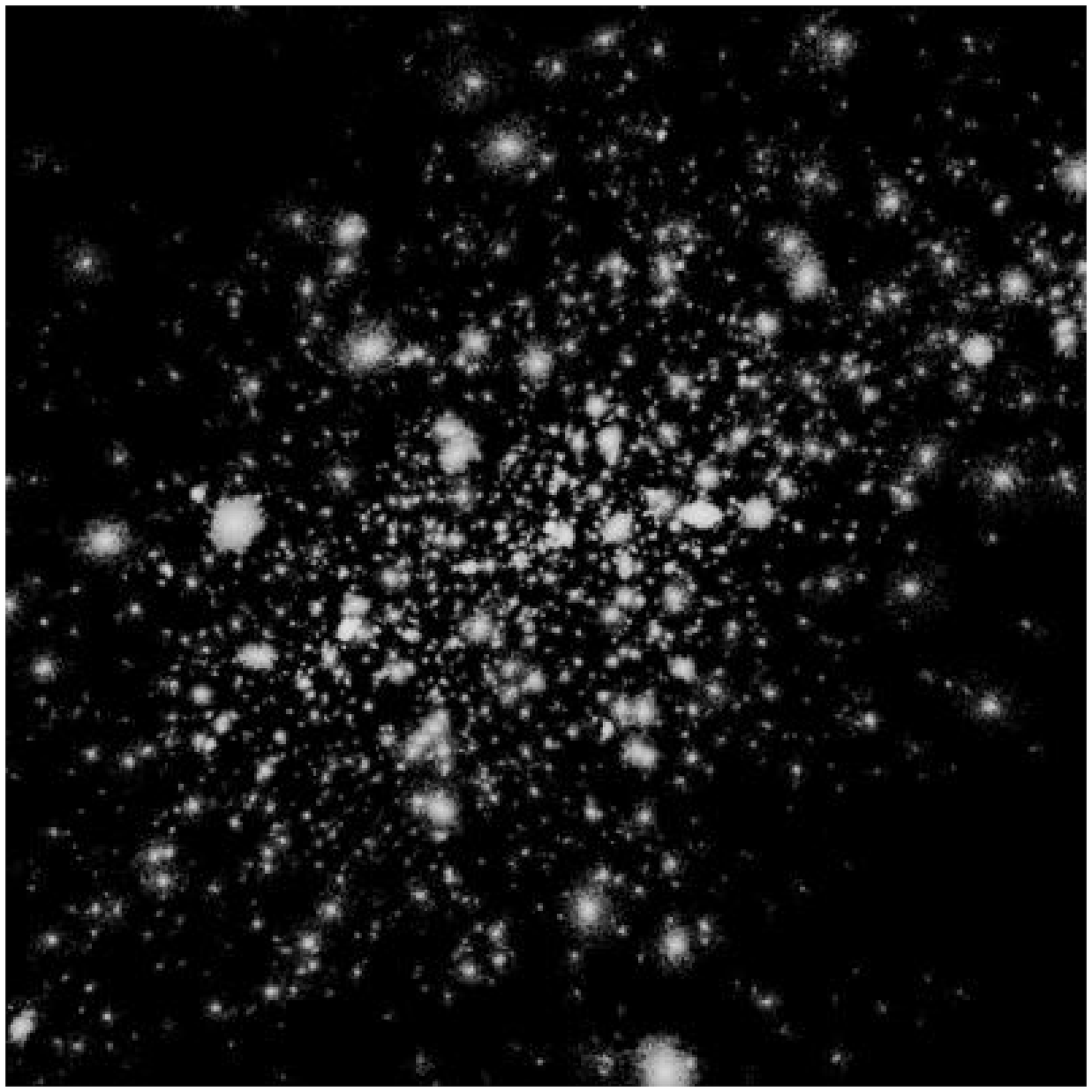}
\end{center}
\caption{Projected image of  halos and detected subhalos.
Top: Results from G1-H, for all particles(left) and detected
subhalos(right). Bottom: Same as the top, but for C1-H.
Box size is $2R_{200}$ for all frames.}
\label{fig:image}
\end{figure*}

Table \ref{tab:prop} lists virial radii($R_{200}$),  masses 
($M_{200}$) and numbers of the particles ($N_{200}$) inside  $R_{200}$, and
circular velocities($V_c$) of the main halo.  Here, $R_{200}$ is the radius
in which the average density is 200 times the critical density.
 Figure \ref{fig:image} shows snapshots  of a galaxy and a cluster
halos from high resolution runs.

\subsection{Method for subhalo detection}

To detect subhalos in simulated galaxy halos, we developed a subhalo
detection program based on the Hierarchical Friends of Friends (HFoF)
method \citep{1999ApJ...516..530K}.  The HFoF method is an extension
of the standard FoF.  The FoF algorithm
with the linking length $h$ detects the regions with the density
higher than $\rho_h=2m/(\frac{4}{3}\pi h^3)$.  In the HFoF method,
FoF is repeatedly  applied to the regions found by FoF, with
decreasing linking length, so 
that we can identify all density peaks. Initially, at level 1, we use
the linking length 
\begin{equation}
h_1 = \left( \frac{1}{2 \bar{\rho}_{crit}} \frac{2m}{\frac{4}{3}\pi}
\right)^{\frac{1}{3}}.
\end{equation}
Here $\bar{\rho}_{crit}$ is the critical density of the Universe
and defined as
$\bar{\rho}_{crit}=3H_0/8\pi G$.

For halos detected at level $i-1$, we apply the FoF method at level $i$ with
linking length 
\begin{equation}
h_i =  2^{-(i-1)/3} h_1,
\end{equation}
i.e., we shrink the linking length by $2^{-1/3}$ at each level. This
factor must be small enough so that the result does not depend on its
value. When FoF of level $i$ is applied to a halo found in level
$i-1$, there are three possibilities. The first one is that a single
halo is again found, but with a smaller number of particles. Second is
that nothing is found, or there are too few particles in the detected
halo. The third possibility is that multiple halos are found.  When
one or more halos are found, we continue to level $i+1$. When the halo
vanished at level $i$, we regard that a single halo is detected at level
$i-1$. The lower limit for the number of particles in a subhalo is
10. 

Any method for subhalo detection based on the isodensity contour has a
problem that many groups found are not gravitationally bound.
To select gravitationally bound groups, we employed the evaporative
method \citep{1997CACR..150}.  Using this method, we can
discard unbound groups. In addition, we can remove unbound particles
from bound groups.  Since the iterative calculation of gravitational potential
necessary for this method is computationally expensive, we used GRAPE-6
to calculate the potential energy in order to accelerate calculation.
We applied this method to all subhalo candidates, even when the number of
particles is very large. Thus, for all halos, we can accurately determine
the bound mass. 

To summarize, our halo finding algorithm is following.  First, the
HFoF method is
applied to all particles to construct density hierarchy.  Then we apply
evaporative method to each of detected density peaks (at the highest
level for which that halo is found).  If the peak is rejected by the
evaporative method, that peak is regarded as not detected and we apply
the same procedure to its parent halo (halo in one level lower).
If the parent halo of one halo contain no other halo, we also regard
that it is detected only at level $i-1$. We apply this procedure to
all hierarchies of subhalos, and regard all remained halos as
detected.

Total numbers of detected subhalos $N_{sub}$ are listed in table
\ref{tab:prop}.  Right panels of figure \ref{fig:image} show the
detected subhalos.

\section{Results}\label{sec:res}

\subsection{Cumulative distribution of subhalos}\label{sec:r1}

\begin{figure*}[t]
   \FigureFile(0.45\hsize,0.45\hsize){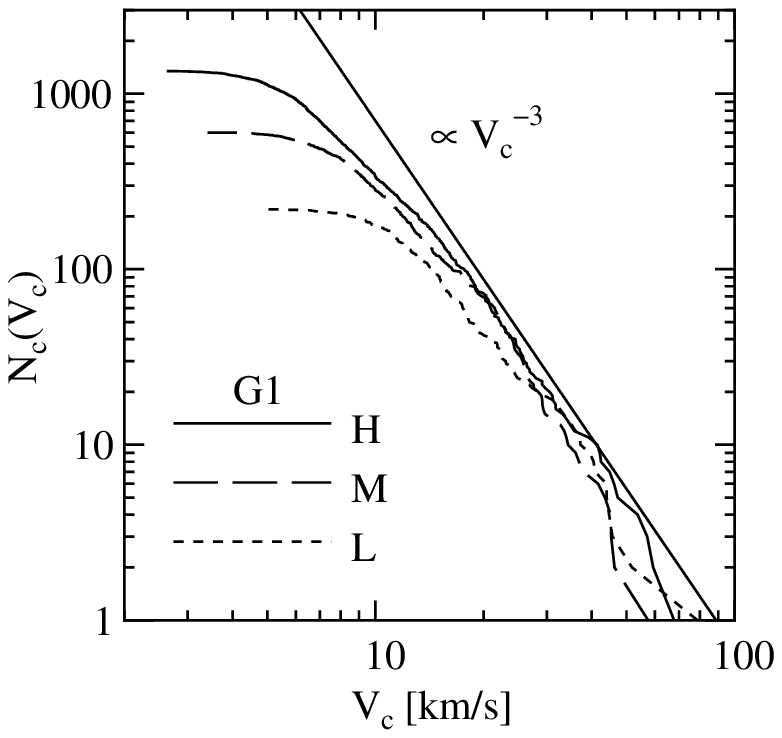}
   \FigureFile(0.45\hsize,0.45\hsize){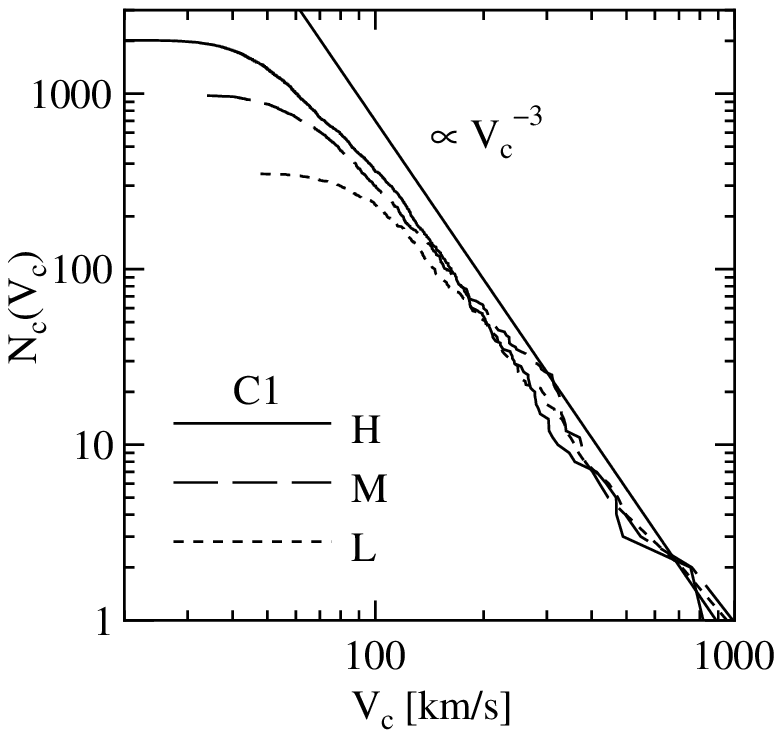}\\
   \FigureFile(0.45\hsize,0.45\hsize){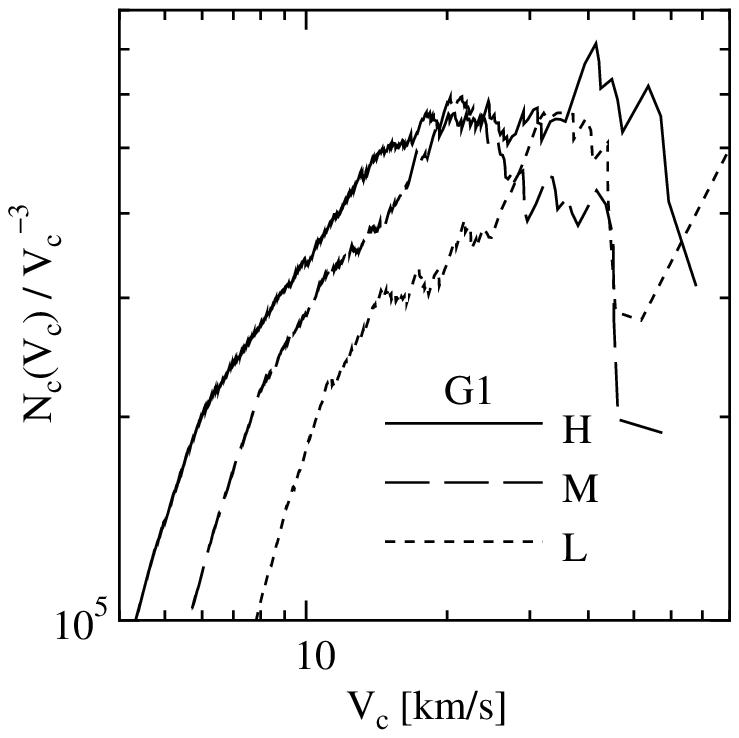}
   \FigureFile(0.45\hsize,0.45\hsize){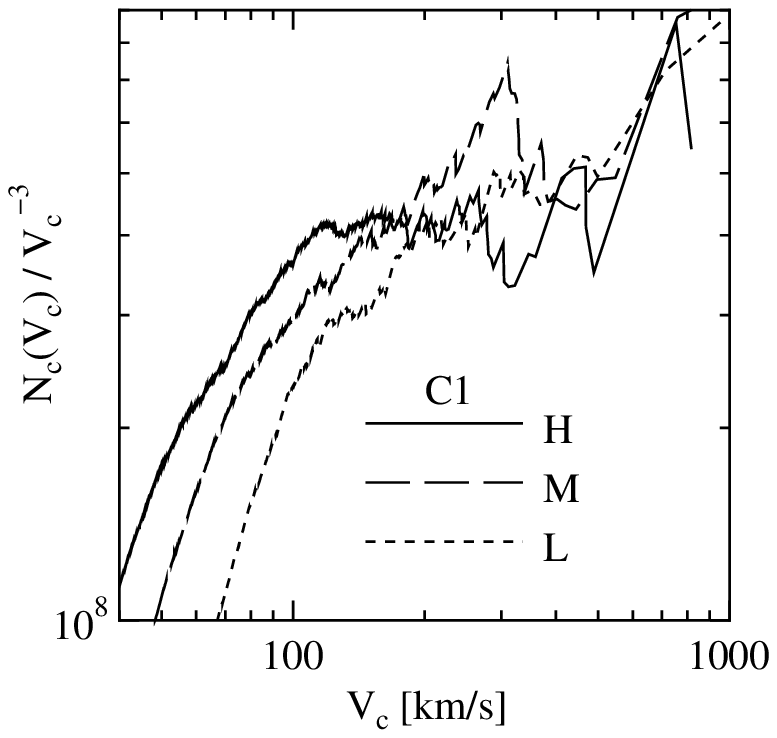}
   \caption{Top: Cumulative number of subhalos as a function of $V_c$.
   Bottom: Cumulative number of $V_c$ normalized by $V_c^{-3}$.
   Left and right are for G1 halos and C1 halos.
  Solid, dashed, and dotted curves indicate the results of
  high, medium, and low resolution runs, respectively.}
  \label{fig:num_vc}
\end{figure*}

\begin{figure*}[t]
   \FigureFile(0.45\hsize,0.45\hsize){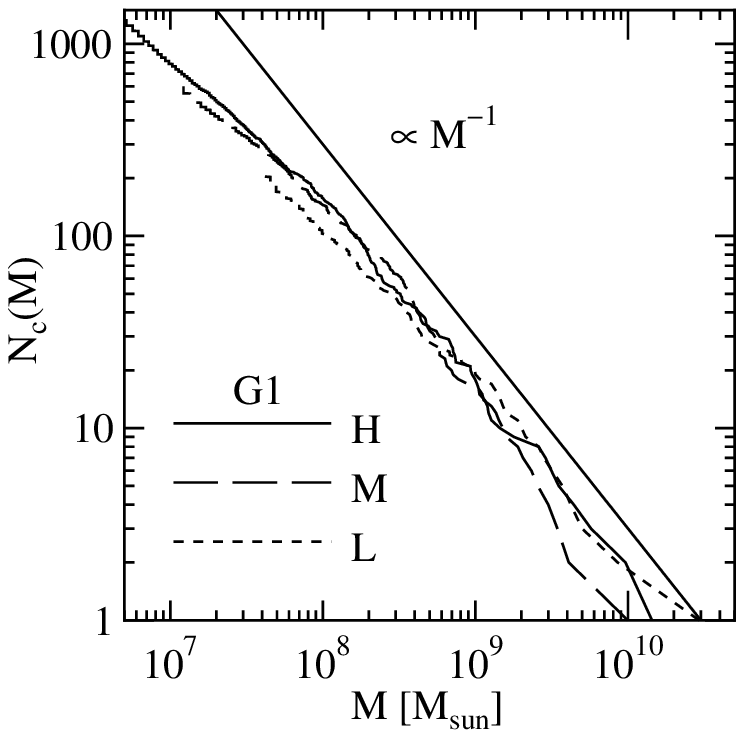}
   \FigureFile(0.45\hsize,0.45\hsize){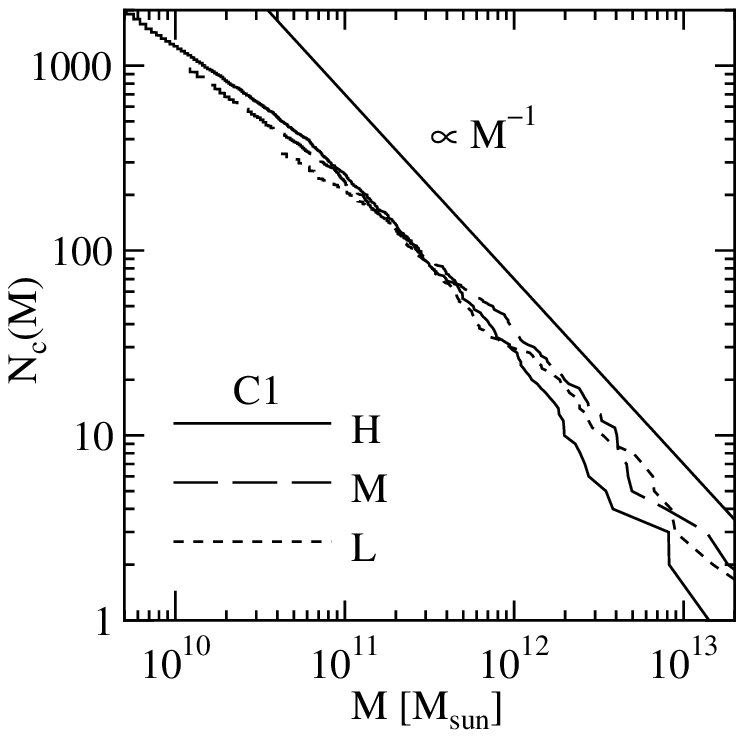}\\
   \FigureFile(0.45\hsize,0.45\hsize){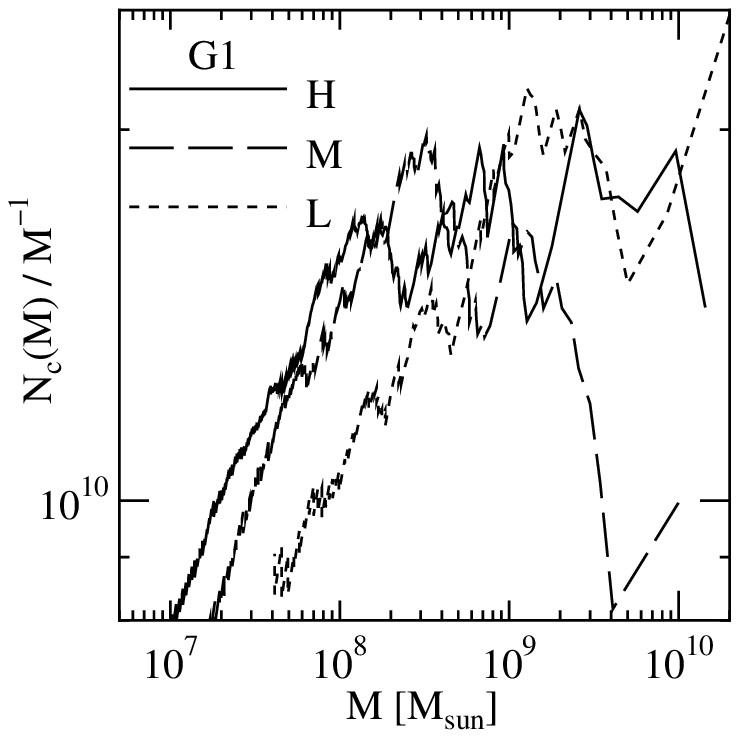}
   \FigureFile(0.45\hsize,0.45\hsize){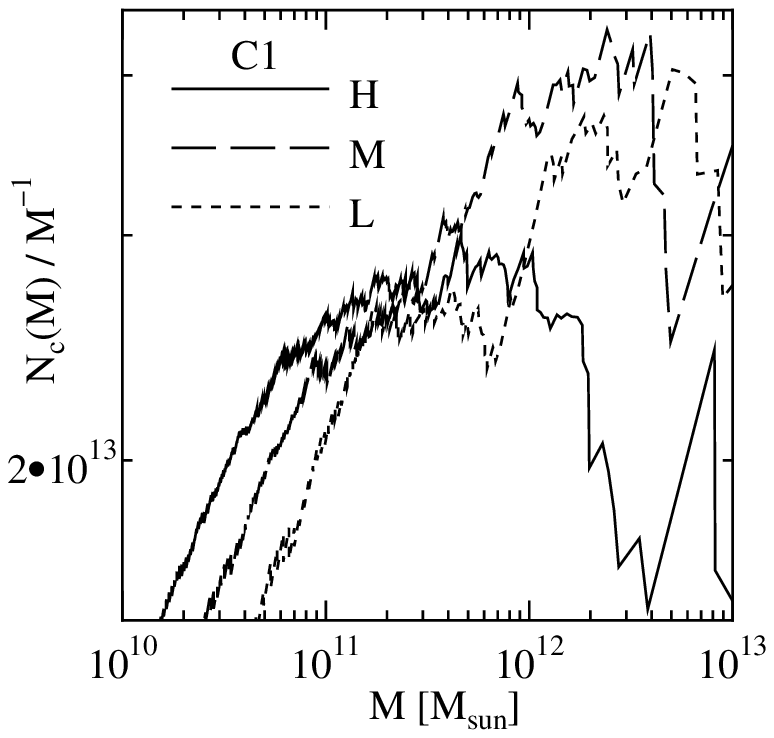}
   \caption{Top: Cumulative number of subhalos as a function of $M$.
   Bottom: Cumulative number of $M$ normalized by $M^{-1}$.
   Left and right of both are for G1 halos and C1 halos.
   Meanings of the curves are the same as in figure \ref{fig:num_vc}.}
   \label{fig:num_mass}
\end{figure*}

Top two panels of figure \ref{fig:num_vc} show the cumulative number
$N_c(V_c)$ of subhalos as the function of the circular velocity $V_c$
for G1 and C1 halos. In both cases, $N_c$ is proportional to $V_c^{-3}$
for large subhalos (higher $V_c$), but for small subhalos the number of
subhalos drops off the power-law line. If we investigate the curves
carefully, we can see that $N_c$ drops off from the power law in two
stages. For example, the curve for G1-H starts to deviate from the
power-law line at $V_c\sim 20{\rm km/s}$, and then show the second
leveling off at $V_c\sim 6{\rm km/s}$. All other curves show similar
two-stage behavior.

To see this tendency more clearly, in bottom panels of figure
\ref{fig:num_vc} we plot the cumulative number normalized by
$V_c^{-3}$. In G1 runs, runs H and M show narrow flat region, where
the relation $N_c\propto V_c^{-3}$ holds. The L (low-resolution) run
shows hardly any flat region. Similar tendency is visible for C1 runs,
though somewhat less clear.

In these panels, it is clear not only the second leveling-offs but
also the first deviations are dependent on the resolution. In other
words, first deviations are numerical artifact, and simulation results
are reliable only for the subhalos larger than these first bendings.

This result indicates that the region where the numerical result is
reliable is quite narrow. For G1-H run, the numerical result is
reliable only for $V_c > 15{\rm km/s}$,  for which only $\sim 100$
halos exist.

Figure \ref{fig:num_mass} shows the same cumulative number of
subhalos, but now as the function of bound mass of halos $M$. We can
see that the result is quite similar to that of figure
\ref{fig:num_vc}, except that the second leveling-off is not visible.

The reason why the leveling-off is visible for velocity and not for
mass is simple. Since we have posed minimum number of particles for
subhalos ($N>10$), there is no subhalo with $N<9$, which is the reason
why the curve terminate suddenly for $N_c(M)$ plot. In the case of
$N_c(V_c)$, halos with $N\ge 10$ but with small $V_c$ were counted,
but halos with $N< 10$ are not counted no matter how high its $V_c$
is. Thus, $N_c$ levels off gradually.

The bottom two panels of figure \ref{fig:num_mass} show the cumulative
number normalized by $M^{-1}$. Again, the bending is clearly
visible. The ``flat'' region is not very clear, perhaps simply due to
small number statistics. For example, if we interpret the result of
run C1-M naively, we might conclude that normalized plot in the mass
range $10^{11}$ to $10^{12}$ is not flat. However, since there are
only 20 subhalos with mass larger than $10^{12}$, this result is not
statistically significant. 

For the small-mass region where runs with different mass resolutions
show different results, the statistical noise is small because the
number of halos is large. At the first sight the result in this region
might look reliable, simply because of this low noise.  However, clear
difference between runs with different resolutions indicate that
result in this region is numerical artifact and is not reliable at
all.

From figure \ref{fig:num_mass} we can estimate the minimum number of
particles in the subhalos above which the number count is
reliable. For run C1-H, the mass at which the normalized cumulative
count bends off from the flat line is between $5\times 10^{10}
\MO$ and $10^{11} \MO$, which correspond to the number of particles in the
subhalos between 100 and 200. For other five runs, bending points are
all similar, between 100 and 200. So we can conclude that the cumulative
count is only reliable for subhalos with number of particles larger
than 200.

\begin{figure}[t]
\FigureFile(0.9\hsize,0.9\hsize){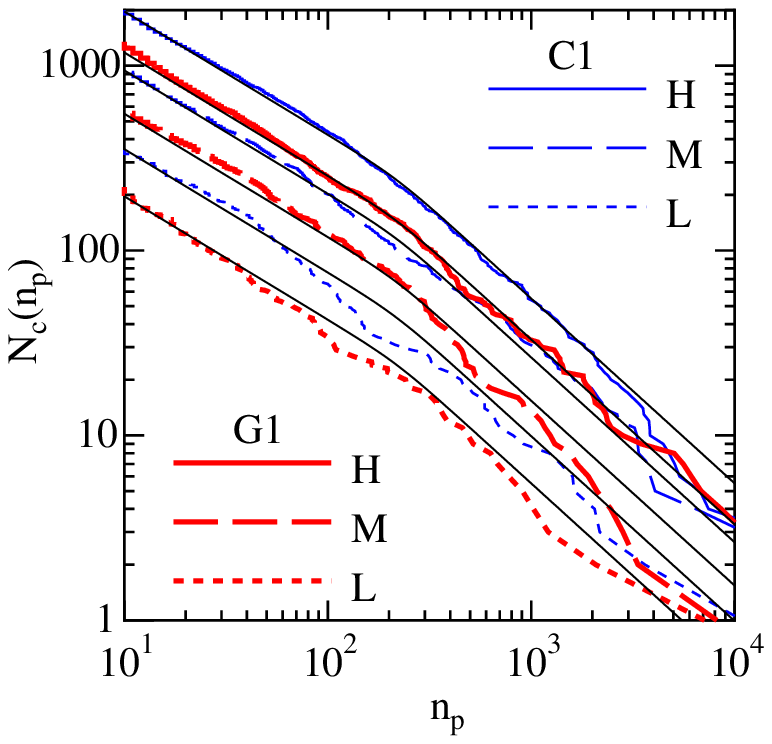}
\caption{Cumulative number of subhalos as a function of
$n_p$, with the fitting function of eq. (\ref{eq:ncfitt}).
Bold curves are in G1 halos and thin curves are in C1 halos.
Line types indicate the resolution of runs, the same as figure \ref{fig:num_vc}
}
\label{fig:num_pnum}
\end{figure}

In order to see this tendency more clearly, in figure
\ref{fig:num_pnum} we show the cumulative number of subhalos plotted
as functions of the number of particles in subhalos. Thin solid curves
are fitting function of the form

\begin{equation}
N_c(n_p)=\frac{C}{\left(\frac{n_p}{n_0}\right)^{\frac{2}{3}}
\left(1+\left(\frac{n_p}{n_0}\right)^5\right)^{\frac{1}{15}}},
\label{eq:ncfitt}
\end{equation}
where $C$ and $n_0$ are fitting parameters. For all curves, we use
$n_0=220$. The other parameter $C$ was chosen by eye.
We can see that for all runs formula (\ref{eq:ncfitt}) gives
excellent fit.

\citet{2005MNRAS.359.1537R} performed simulations similar to what we
performed, and measured the cumulative number of subhalos as functions
of circular velocity and mass. They have not reported this bending at
$n_p=200$. For velocity distribution, they did not see this bending
simply because their cutoff number of particles for subhalos is much
bigger than what we used. As the result, the second leveling off
occurs almost simultaneously as the first bending. For mass
distribution function, their figure 11 shows the same bending
behavior as we see in figure \ref{fig:num_pnum}, though the tendency
is less clear simply because of the way the data are plotted, and they
did not discuss the shape of the mass distribution function.
Qualitatively same tendencies were reported by
\citet{2004MNRAS.355..819G} in galaxy
scale and \citet{2004MNRAS.352..535D} in cluster scale.

\subsection{True number of subhalos and comparison with the observation}

For our largest calculation with 4M particles in the virial radius of
the parent halo, we found that the distribution of subhalos is
reliable only up to 100-200 most massive subhalos. For smaller
subhalos, the dependence of the cumulative number on both the circular
velocity and mass becomes significantly shallower than the simple
power-law. Thus, we should assume that the number of subhalos formed
in real CDM cosmology would show this simple power-law behavior down to
mass or circular velocity much lower than the limit for which the
current simulations can give reliable result. 

The classic work by
M99 used around 1M particles for the parent halo, and
compared the distribution of around 1,000 subhalos. On the other
hand, we can conclude that their result is actually reliable only for
around 20 most massive halos, and the number of smaller subhalos was
significantly underestimated.

\begin{figure}[t]
\begin{center}
\FigureFile(0.9\hsize,0.9\hsize){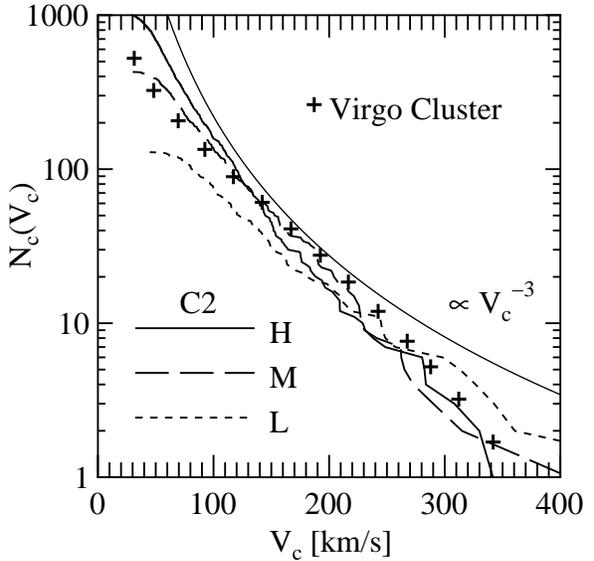}
\end{center}
\caption{Cumulative number of subhalos as a function of $V_c$ in runs C2-X.
Solid, dashed and dotted curves indicate runs H, M, and L, as in figure
\ref{fig:num_vc}. Smooth solid curve shows $V_c^{-3}$.
Crosses are the distribution of galaxies in Virgo Cluster, used in figure 2 of
M99. }
\label{fig:num_vel_c123}
\end{figure}

Figure \ref{fig:num_vel_c123} shows the cumulative number of subhalos
for runs C2-H, M and L, as well as that estimated for the Virgo cluster in
the way same as in M99. We can see that run C2-M, which
used the number of particles similar to that in M99, shows the best
agreement with the observation. The result of run C2-H shows the slope
steeper than the observation starting at $V_c=200{\rm km/s}$.
The actual number count for run C2-H becomes higher than
observation only at $V_c=140{\rm km/s}$ or around, because the number
of massive halos in  run C2-H is slightly smaller than the
observational value, while C2-M shows better fit for
$V_c > 140{\rm km/s}$. This difference between C2-H and C2-M is
purely due to the fluctuation in the initial model.
The high-wave number initial fluctuations we added for
re-simulation are not exactly the same for these runs.

As we
have seen in figure \ref{fig:num_vc}, the number count for this run
drops off from the power-law line also around  $V_c=140{\rm
km/s}$. Thus, the actual number of galaxies observed in the Virgo
cluster goes even lower than the simulation which gives the number
lower than the power-law. For reference, we drew the power-law curve in figure
\ref{fig:num_vel_c123}. The curve implies that, ideally,
for $V_c \sim 100 {\rm km/s}$, the difference
is already significant, and for 50km/s, there is more than a factor of 10
difference. 

\subsection{Distribution of subhalos in ${M - V_c}$ plane}

\begin{figure*}[t]
   \FigureFile(0.45\hsize,0.45\hsize){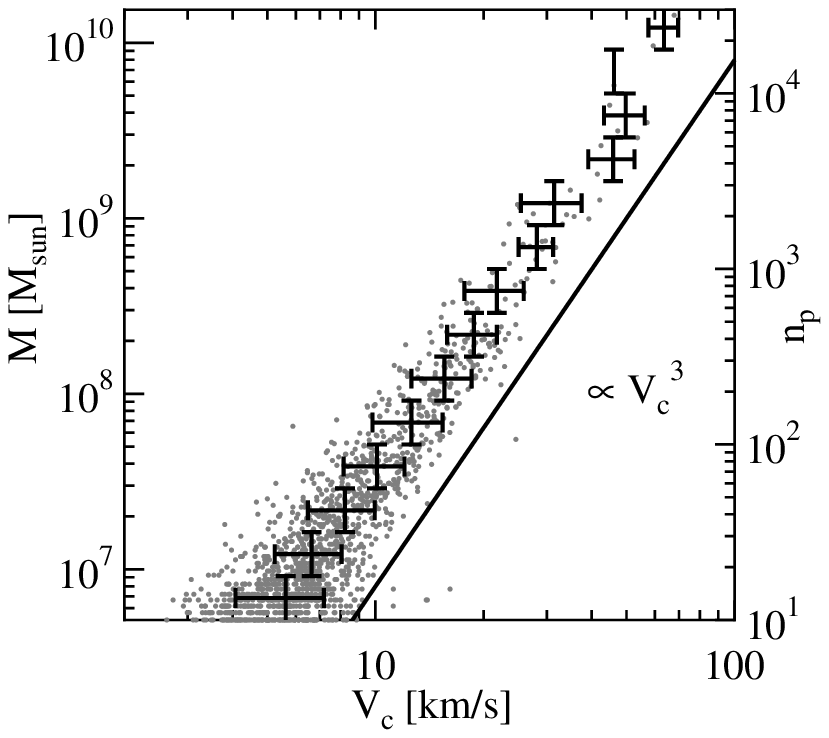}
   \FigureFile(0.45\hsize,0.45\hsize){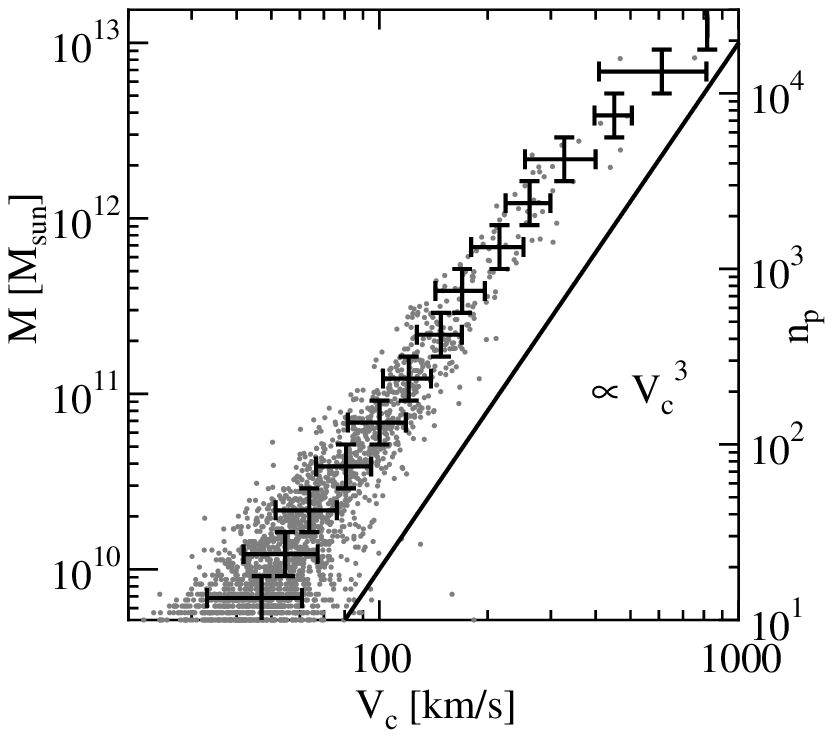}\\
   \caption{Distribution of $V_c$ and $M$ of subhalos.
   Left is in G1-H halo and right is in C1-H halo.
   Horizontal error bars indicate the unbiased dispersion of
   the $V_c$ of subhalos inside the bins of mass, whose range
   is shown by vertical error bars.
   }
   \label{fig:vel_mass}
\end{figure*}

In figure \ref{fig:vel_mass} the mass of each subhalo is plotted against
its circular velocity for runs G1-H and C1-H.
Figure \ref{fig:vel_mass} shows that the correlation
between $M$ and $V_c$ is fairly tight, and well fitted by $M \propto
V_c^3$, for both runs G1 and C1.  This tight relation, however,
becomes somewhat loose for the velocity smaller than 20 km/s in the
case of the galaxy scale G1 and than 100 km/s in the case of the
cluster scale C1. We show the $1\sigma$ unbiased variances  in figure
\ref{fig:vel_mass}, from which we can sort of see that the
distribution of circular velocity for a given mass range is wider for
lower mass.

\begin{figure*}[t]
   \FigureFile(0.45\hsize,0.45\hsize){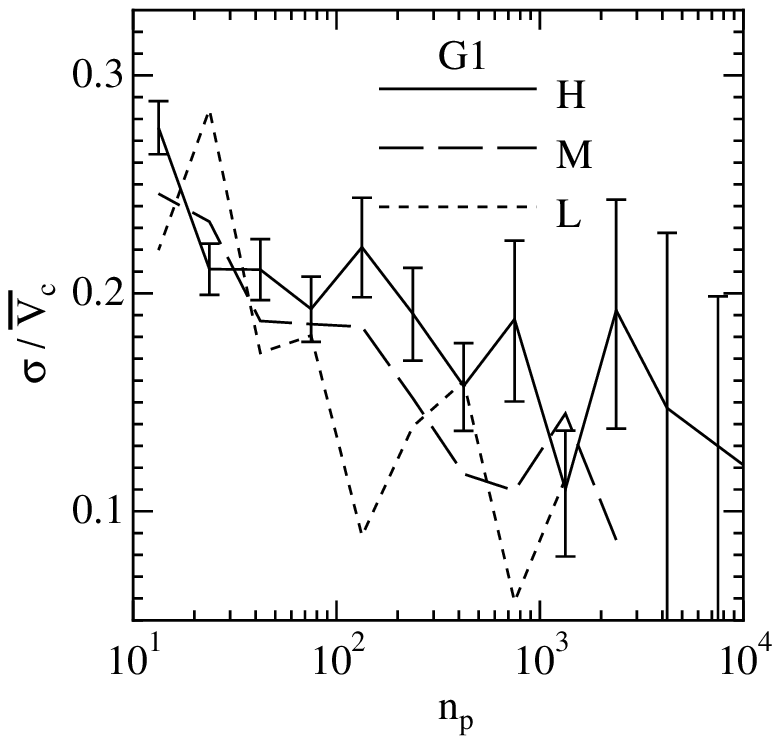}
   \FigureFile(0.45\hsize,0.45\hsize){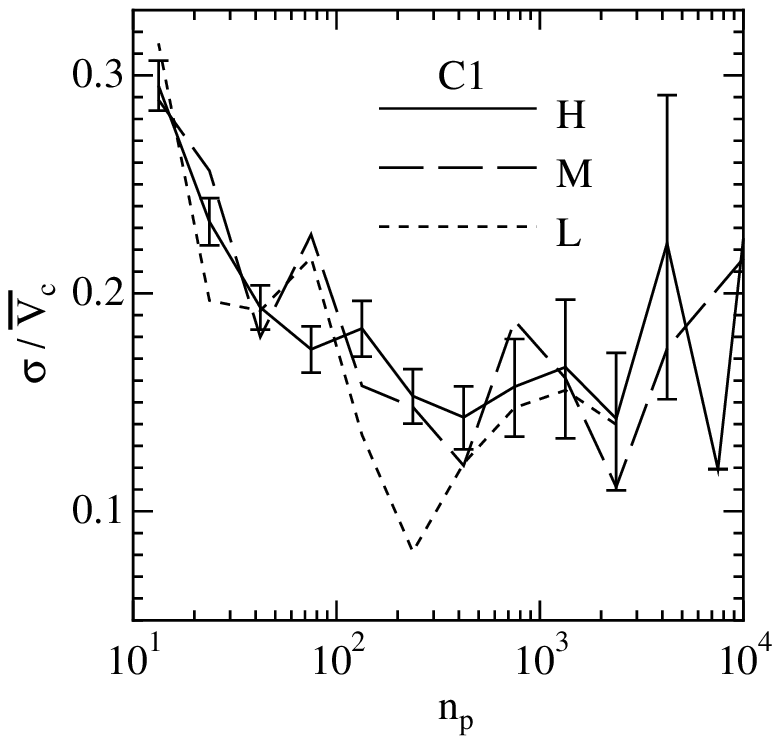}\\
   \FigureFile(0.45\hsize,0.45\hsize){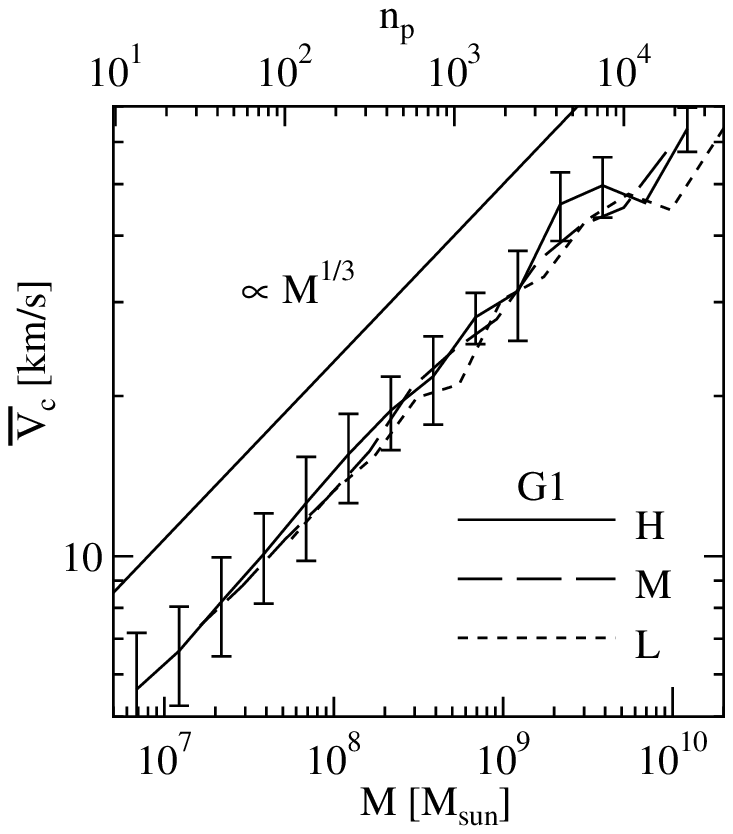}
   \FigureFile(0.45\hsize,0.45\hsize){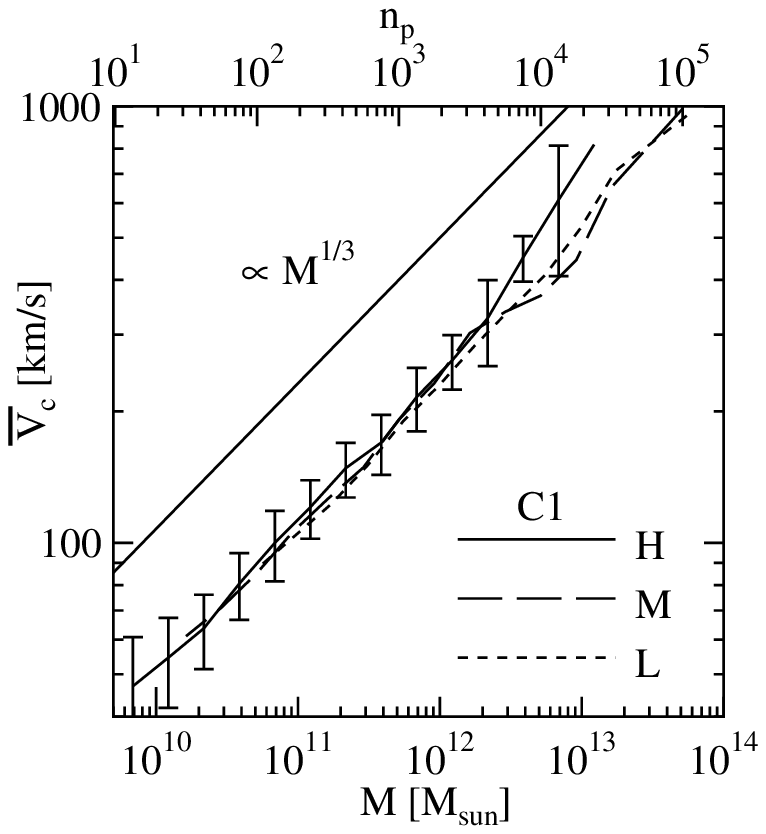}
   \caption{Top: The ratio between the dispersion($\sigma$) and the average
   ($\overline{V_c}$)
   of $V_c$ as a function of $n_p$.
   Vertical error bars are derived from dispersion of
   $\sigma$ and $\overline{V_c}$.
   Bottom: Average of $V_c$ as a function of $M$.
   Upper horizontal axis shows the corresponding
   $n_p$ for high resolution run.
   Vertical error bars indicate the dispersion of $V_c$, the same data
   as plotted in figure \ref{fig:vel_mass}.
   Left and right are for G1 and C1 runs, respectively.
   The meanings of curves are the the same as that in figure \ref{fig:num_vc}.}
   \label{fig:vel_disp_ave}
\end{figure*}

To see this tendency more clearly, in figure \ref{fig:vel_disp_ave} we
plot the unbiased variance of the circular velocity as the function of
halo size. Here, we used the number of particles in subhalos as the
indicator of the size of the halo. We can see that the variance shows
essentially the same behavior for three runs with different mass
resolutions, at least for cluster-scale runs C1-X. This behavior
indicates that the increase of the variance in low-mass halos is not
the physical reality but the numerical artifact. The real variance is
probably around 15\%, independent of the mass of subhalos. 
The galaxy-scale runs show a similar  tendency.

The number of particles below which the variance becomes large is
around 200, the same as the number of particles below which the number
count becomes unreliable. Thus, these deviations are probably driven
by the same mechanism.

\section{Summary and discussions}

\subsection{Distribution of subhalos}

We carried out large scale $N$-body simulations of formation of CDM
halos.  We simulated the formation of both a galactic size halo and a
cluster size halo.  We determined the masses and circular velocities
of subhalos for both cases.  We found that the cumulative distribution of
subhalos, both as a function of the circular velocity and as that of
the mass, is affected by mass resolution of the simulation.  In all
runs, the number of subhalos $n$ with more than 200 particles is well
expressed by a single power-law of $n \propto V_c^{-3}$.  For the
subhalos with less than 200 particles, the number of subhalos becomes
smaller than this power-law line.

We investigated the distribution of subhalos in the plane of their
masses and circular velocities, and found that the there is a tight
correlation with  $M \propto V_c^{3}$. The tightness of the
correlation again depends on the number of particles in subhalos, and
for subhalos with less than 200 particles the correlation becomes
weaker. 

\subsection{``Dwarf galaxy problem'' in clusters of galaxies}

From runs G1-X, the number of subhalos with $n_p>200$ are
about 30 for run G1-L, 200 for run G1-H. So
roughly speaking, our result implies that the cumulative distribution
of subhalos obtained by $N$-body simulation is reliable only for
$N/(2\times 10^4)$ most massive halos, where $N$ is the number of
particles in the virial radius of the parent halo. Thus, results
obtained with, for example, 1M particles is okay for only the first 50
subhalos. This result means that the good agreement of the observation
and simulation result obtained with around 1M particles is actually a
numerical artifact, and there is a serious discrepancy between the
observed number of galaxies in clusters like the Virgo cluster and the number
of subhalos obtained by  $N$-body simulation. Ideally the number of subhalos
with, say, circular velocity of 100km/s is factor two or three times bigger
than the observed count, and for 50km/s this factor is more than 10.

Most ``solutions'' so far proposed for the dwarf galaxy problem
were aimed at reducing the number of dwarfs with $V_c \sim 20{\rm
km/s}$ or less. 
(\cite{2000ApJ...542..622C}, \cite{2001ApJ...547..574D},
 \cite{2002MNRAS.335L..84S}, K04) However, most of them do not work for halos
with mass more than $10^{11}\MO$.

One serious theoretical problem with the subhalos with mass
$10^{11}\MO$ is that when they were first formed they were
generally much more massive. (\cite{2004MNRAS.348..333D}, K04) Most of 
their mass have been stripped out through the tidal interaction with the
parent halo. In other words, they were initially almost as massive as
the halo of our galaxy. Thus, a ``solution'' which can account for the
discrepancy between the observed number of galaxies in clusters and
the subhalos in a CDM halo need to be able to reduce the number of
halos with initial virial mass as large as $10^{12}\MO$ by a
factor of two or so.

\bigskip
{\bf Acknowledgment } We thank Toshiyuki Fukushige and
Hideki Yahagi for helpful discussions. The
simulation was partly done using IBM pSeries 690 of the Data
Reservoir Project of the University of Tokyo.
This research is partially supported by the Special Coordination Fund
for Promoting Science and Technology (GRAPE-DR project),
and a grant for the 21st Century COE program of the University of Tokyo,
``Quantum Extreme Systems and Their Symmetries''
Ministry of Education, Culture, Sports, Science and Technology, Japan.


\end{document}